\documentclass[floats,floatfix,prb,amsmath,showpacs,twocolumn]{revtex4} 

\usepackage{epsfig}
\usepackage{colordvi}
\usepackage{graphicx}
\usepackage{dcolumn}
\def\E{{\cal E}}
\def\Pmac{{\bf P}_{\rm mac}}
\def\o{\omega}
\def\q{{\bf q}}
\def\k{{\bf k}}
\def\g{{\bf g}}
\def\R{{\bf R}}
\def\rr{{\bf r}}

\begin{document}

\title{First-Principles Perturbative Computation of Phonon Properties
of Insulators\\ in Finite Electric Fields}

\author{Xinjie Wang and David Vanderbilt}
\affiliation{Department of Physics and Astronomy, Rutgers University,
	Piscataway, NJ 08854-8019, USA}
\date{\today}

\begin{abstract}
We present a perturbative method for calculating phonon properties
of an insulator in the presence of a finite electric field. The starting 
point is a variational total-energy functional with a field-coupling 
term that represents the effect of the electric field.
This total-energy functional is expanded in small atomic 
displacements within the framework of density-functional perturbation 
theory. The linear response of field-polarized Bloch functions to atomic 
displacements is obtained by minimizing the second-order derivatives 
of the total-energy functional.  In the general case of nonzero phonon
wavevector, there is a subtle interplay between the couplings between
neighboring k-points introduced by the presence of the electric field
in the reference state, and further-neighbor k-point couplings
determined by the wavevector of the phonon perturbation.  As a
result, terms arise in the perturbation expansion that take the
form of four-sided loops in k-space.
We implement the method in the {\tt ABINIT} code and perform illustrative 
calculations of the field-dependent phonon frequencies for
III-V semiconductors. 
\end{abstract}


\vskip9pc

\maketitle

\section{Introduction}
\label{sec:intro}

The understanding of ferroelectric and piezoelectric materials,
whose physics is dominated by soft phonon modes, has benefited
greatly from the availability of first-principles methods for
calculating phonon properties. In general, these methods can be
classified into two main types, the direct or frozen-phonon
approach\cite{direct1,direct2} and the linear-response
approach.\cite{dfpt3,dfpt2} In the former approach, the properties
of phonons at commensurate wavevectors are obtained from supercell
calculations of forces or total-energy changes between
between equilibrium and distorted structures.  In the latter
approach, based on density-functional perturbation theory (DFPT),
expressions are derived for the second derivatives of the total energy
with respect to atomic displacements, and these are calculated by solving
a Sternheimer equation\cite{dfpt3} or by using minimization
methods.\cite{dfpt2,dfpt1}
Compared to the direct approach, the linear-response
approach has important advantages in that time-consuming supercell
calculations are avoided and phonons of arbitrary wavevector can
be treated with a cost that is independent of wavevector.
However, existing linear-response methods work only
at zero electric field.

The development of first-principles methods for treating the effect
of an electric field $\E$ in a periodic system has been impeded by
the presence of the electrostatic potential $\E \cdot {\bf r}$ in the
Hamiltonian.
This potential is linear in real space and unbounded from
below, and thus is incompatible with periodic boundary conditions. The
electronic bandstructure becomes ill-defined after application of a
potential of this kind.  Many attempts have been made to overcome this
difficulty.  For example, linear-response approaches have been used
to treat the electric field as a perturbation.\cite{dfpt1,baroni01}
It is possible to formulate these approaches so that only the off-diagonal
elements of the position operator, which remain well defined, are needed,
thus allowing for the calculation of Born effective charges,
dielectric constants, etc. Since it is a perturbative approach, a
finite electric field cannot be introduced.

Recently, a total-energy method for treating insulators in nonzero
electric fields has been proposed.\cite{souza02,umari02} In this
approach, an electric enthalpy functional is defined as a sum of the usual
Kohn-Sham energy and an $\E\cdot{\bf P}$ term expressing the linear
coupling of the electric field to the polarization $\bf P$.
The enthalpy functional is minimized with
respect to field-polarized Bloch states, and the information on the
response to the electric field is contained in these optimized Bloch
states.  Using this approach, it is possible to carry out calculations
of dynamical effective charges, dielectric susceptibilities,
piezoelectric constants, etc., using finite-difference
methods.\cite{souza02,umari02}  It would also be possible to use it
to study phonon properties in finite electric field, but with
the aforementioned limitations (large supercells, commensurate
wavevectors) of the direct approach.

In this work, we build upon these recent developments by showing how
to extend the linear-response methods so that they can be applied to
the finite-field case.  That is, we formulate DFPT for the case in
which the unperturbed system is an insulator in a finite electric
field.  Focusing on the case of phonon perturbations, we derive
a tractable computational scheme and demonstrate its effectiveness
by carrying out calculations of phonon properties of polar semiconductors
in finite electric fields.

This paper is organized as follows.  In Sec.~\ref{sec:tef} we review the
total-energy functional appropriate for describing an insulator
in an electric field, and discuss the effect of the electric
field on the phonon frequencies both for our exact theory and for a
previous approximate theory. The second-order expansion of the
total-energy functional is derived in Sec.~\ref{sec:peeef}, and
expressions for the force-constant matrix are given, first for
phonons at the Brillouin zone center and then for arbitrary
phonons.  In Sec.~\ref{sec:test} we report some test calculations
of field-induced changes of phonon frequencies in the III-V
semiconductors GaAs and AlAs. Section V contains a brief summary
and conclusion.

\section{Background and definitions}
\label{sec:tef}

\subsection{Electrical enthalpy functional}

We start from the electric enthalpy functional \cite{souza02}
\begin{equation}
F[\R;\psi;\E]=E_{\rm KS}[\R;\psi]
-\Omega \E\cdot \Pmac[\psi]  ,
\label{eq:Ftotal}
\end{equation}
where $E_{\rm KS}$ has the same form as the usual Kohn-Sham energy
functional in the absence of an electric field. Here $\Omega$
is the cell volume, $\Pmac$ is the macroscopic
polarization, $\E$ is the
homogeneous electric field, $\R$ are the atomic positions, and
$\psi$ are the field-polarized Bloch functions. Note that $\Pmac$
has both ionic and electronic contributions. The former is an
explicit function of $\R$, while the latter is an implicit function
of $\R$ through the Bloch functions, which also
depend on the atomic positions.  When an electric field is present,
a local minimum of this functional describes a long-lived
metastable state of the system rather than a true ground state
(indeed, a true ground state does not exist in finite electric
field).\cite{souza02}

According to the modern theory of polarization,\cite{smith93}
the electronic contribution to the macroscopic polarization
is given by
\begin{equation}
{\bf P}_{\rm mac}=\frac{ief}{(2\pi)^3}\sum_{n=1}^{M} \int_{\rm BZ}d\k
  \langle u_{\k n}| \nabla_k |u_{\k n}\rangle ,
\label{eq:Pmacc}
\end{equation}
where $e$ is the charge of an electron ($e<0$), $f$=2 for spin
degeneracy, $M$ is the number of occupied bands, $u_{\k n}$ are
the cell-periodic Bloch functions, and the integral is over the
Brillouin zone (BZ).  Making the transition to
a discretized k-point mesh, this can be written in a form
\begin{equation}
{\bf P}_{\rm mac}=\frac{ef}{2\pi\Omega}\sum_{i=1}^{3}\frac{{\bf a}_i} {N_{\perp }^{(i)}}
\sum_{l=1}^{N_{\perp}^{(i)}}\textrm{Im}
     \ln \prod_{j=1}^{N_i}\det S_{\k_{lj},\k_{l,j+1}}
\label{eq:Pmac}
\end{equation}
that is amenable to practical calculations.
In this expression, for each lattice direction $i$
associated with primitive lattice vector ${\bf a}_i$, the BZ
is sampled by $N_\perp^{(i)}$ strings of k-points, each with
$N_i$ points spanning along the reciprocal lattice vector conjugate
to ${\bf a}_i$, and
\begin{equation}
(S_{\k\k'})_{mn}= \langle u_{m\k}|u_{n\k'}\rangle
\label{eq:Sdef}
\end{equation}
are the overlap matrices between cell-periodic
Bloch vectors at neighboring locations along the string.
Because Eqs.~(\ref{eq:Pmacc}-\ref{eq:Pmac}) can be
expressed in terms of Berry phases, this is sometimes referred to as
the ``Berry-phase theory'' of polarization.

\subsection{Effect of electric field on phonon frequencies}
\label{sec:efieldonfreq}

\subsubsection{Exact theory}
\label{sec:exact}

We work in the framework of a classical zero-tem\-per\-a\-ture
theory of lattice dynamics, so that quantum zero-point and thermal
anharmonic effects are neglected.  In this context, the phonon
frequencies of a crystalline insulator depend upon an applied electric
field in three ways:
(i) via the variation of the equilibrium lattice vectors (i.e., strain)
with applied field; (ii) via the changes in the equilibrium atomic
coordinates, even at fixed strain; and (iii) via the changes in the
electronic wavefunctions, even at fixed atomic coordinates and strain.
Effects of type (i) (essentially, piezoelectric and electrostrictive
effects) are beyond the scope of the present work, but are relatively
easy to include if needed.  This can be done by computing the relaxed
strain state as a function of electric field using the approach of
Ref.~\onlinecite{souza02}, and then computing the phonon frequencies
in finite electric field for these relaxed structures using the methods
given here.  Therefore, in the remainder of the paper, the lattice vectors
are assumed to be independent of electric field unless otherwise stated,
and we will focus on effects of type (ii) (``lattice effects'') and
type (iii) (``electronic effects'').

In order to separate these two types of effects, we first write the
change in phonon frequency resulting from the application of the
electric field as
\begin{equation}
\Delta\o(\q;\E) = \o(\q;\R_\E,\E) - \o(\q;\R_0,0) ,
\label{eq:dwtot}
\end{equation}
where $\o(\q;\R,\E)$ is the phonon frequency extracted from the
second derivative of the total energy of Eq.~(\ref{eq:Ftotal}) with
respect to the phonon amplitude of the mode of wavevector $\q$, evaluated
at displaced coordinated $\bf R$ and with electrons experiencing electric
field $\E$.  Also, $\R_\E$ are the relaxed atomic coordinates at
electric field $\E$, while $\R_0$ are the relaxed atomic coordinates
at zero electric field.  Then Eq.~(\ref{eq:dwtot}) can be decomposed as
\begin{equation}
\Delta\o(\q;\E) = \Delta\o_{\rm el}(\q;\E) + \Delta\o_{\rm ion}(\q;\E)
\label{eq:dwtwo}
\end{equation}
where the electronic part of the response is defined to be
\begin{equation} 
\Delta\o_{\rm el}(\q;\E) = \o({\q;\bf R}_0,\E) - \o(\q;\R_0,0)
\label{eq:wel}
\end{equation}
and the lattice (or ``ionic'')  part of the response is defined to be
\begin{equation}
\Delta\o_{\rm ion}(\q;\E) = \o(\q;\R_\E,\E) - \o(\q;\R_0,\E) .
\label{eq:wion}
\end{equation}
In other words, the electronic contribution reflects the influence
of the electric field on the wavefunctions, and thereby
on the force-constant matrix, but evaluated at the zero-field
equilibrium coordinates.  By contrast, the ionic contribution
reflects the additional frequency shift that results from the
field-induced ionic displacements.

The finite-electric-field approach of
Refs.~\onlinecite{souza02}-\onlinecite{umari02}
provides the methodology needed to compute the relaxed coordinates
$\R_\E$, and the electronic states, at finite electric field $\E$.
The remainder of this work is devoted to developing and testing the
techniques for computing $\o(\q;\R,\E)$ for given $\q$, $\rm R$,
and $\E$, needed for the evaluation of Eq.~(\ref{eq:dwtot}).  We shall
also use these methods to calculate the various quantities needed to
perform the decomposition of Eqs.~(\ref{eq:dwtwo}-\ref{eq:wion}), so
that we can also present results for $\Delta\o_{\rm el}$ and
$\Delta\o_{\rm ion}$ separately in Sec.~\ref{sec:test}.

\subsubsection{Approximate theory}
\label{sec:approx}

Our approach above is essentially an exact one, in which Eq.~(\ref{eq:dwtot})
is evaluated by computing all needed quantities at finite electric
field.  However, we will also compare our approach with an approximate
scheme that has been developed in the literature over the last few
years,\cite{sai02,fu03,naumov05,antons05} in which the electronic
contribution is neglected and the lattice contribution is approximated
in such a way that the finite-electric-field approach of
Refs.~\onlinecite{souza02}-\onlinecite{umari02} is not needed.

This approximate theory can be formulated by starting with the
approximate electric enthalpy functional\cite{sai02}
\begin{equation}
F[\R;\E]=E^{(0)}_{\rm KS}[\R]
-\Omega \E\cdot \Pmac^{(0)}[\R] 
\label{eq:Fsai},
\end{equation}
where $E^{(0)}_{\rm KS}[\R]$ is the {\it zero-field} ground-state
Kohn-Sham energy at coordinates $\bf R$, and $\Pmac^{(0)}$ is the
corresponding {\it zero-field} electronic polarization.  In the
presence of an applied electric field $\E$, the equilibrium coordinates
that minimize Eq.~(\ref{eq:Fsai}) satisfy the force-balance equation
\begin{equation}
-\frac{dE^{(0)}_{\rm KS}}{d\R} +{\bf Z}^{(0)}\cdot\E=0
\label{eq:forcebal}
\end{equation}
where ${\bf Z}^{(0)}=\Omega\, d\Pmac^{(0)}/d\R$ is the {\it zero-field}
dynamical effective charge tensor.  That is, the sole effect of the
electric field is to make an extra contribution to the atomic forces
that determine the relaxed displacements; the electrons themselves do
not ``feel'' the electric field except indirectly through these
displacements.  In Ref.~\onlinecite{sai02}, it was shown that this
theory amounts to treating the coupling of the electric field to the
electronic degrees of freedom in linear order only, while treating the
coupling to the lattice degrees of freedom to all orders.
Such a theory has been shown to give good accuracy in cases
where the polarization is dominated by soft polar phonon modes, but not
in systems in which the electronic and lattice polarizations are
comparable.\cite{sai02,fu03,naumov05,antons05,dieguez06}

In this approximate theory, the effect of the electric field on the
lattice dielectric
properties\cite{antons05} and phonon frequencies\cite{naumov05}
comes about through the field-induced atomic displacements.  Thus, in the
notation of Eqs.~(\ref{eq:dwtot}-\ref{eq:wion}), the frequency shift (relative
to zero field) is
\begin{equation}
\Delta\o_{\rm ion}'(\q;\E) = \o(\q;\R'_\E,0) - \o(\q;\R_0,0)
\label{eq:wfu}
\end{equation}
in this approximation, where $\R'_\E$ is the equilibrium
position according to Eq.~(\ref{eq:forcebal}).
We will make comparisons between the exact $\R_\E$ and the
approximate $\R'_\E$,
and the corresponding frequency shifts $\Delta\o_{\rm ion}(\q,\E)$
and $\Delta\o_{\rm ion}'(\q,\E)$ later in Sec.~\ref{sec:test}.

\section{Perturbation expansion of the electric enthalpy functional}
\label{sec:peeef}

We consider an expansion of the properties of the system in terms
of small displacements $\lambda$ of the atoms away from their
equilibrium positions, resulting in changes in the charge
density, wavefunctions, total energy, etc. We will be more
precise about the definition of $\lambda$ shortly.  We adopt a notation
in which the perturbed physical quantities are expanded in powers
of $\lambda$ as
\begin{equation}
Q(\lambda)=Q^{(0)}+\lambda Q^{(1)}+\lambda^2 Q^{(2)}+\lambda^3 Q^{(3)}+ ...
\end{equation}
where $Q^{(n)}=(1/n!)d^nQ/d\lambda^n$.  The immediate dependence
upon atomic coordinates is through the external potential
$v_{{\rm ext}}(\lambda)$, which has no electric-field dependence and
thus depends upon coordinates and pseudopotentials in the same way
as in the zero-field case.  The changes in electronic wavefunctions,
charge density, etc. can then be regarded as being induced by the
changes in $v_{{\rm ext}}$.

\subsection{Zero q wavevector case}
\label{sec:zeroq}

The nuclear positions can be expressed as 
\begin{equation}
\R_{n \nu} = {\bf t}_n+{\bf d}_{\nu}+{\bf b}_{n \nu} ,
\end{equation}
where ${\bf t}_n$ is a lattice vector, ${\bf d}_{\nu}$ is
a basis vector within the unit cell, and ${\bf b}_{n \nu}$
is the instantaneous displacement of atom $\nu$ in cell $n$.
We consider in this section a phonon of wavevector $\q={\bf 0}$,
so that the perturbation does not change the periodicity of the
crystal, and the perturbed wavefunctions satisfy the same periodic
boundary condition as the unperturbed ones.  To be more precise,
we choose one sublattice $\nu$ and one Cartesian direction $\alpha$
and let $b_{n\nu\alpha}=\lambda$ (independent of $n$), so that we
are effectively moving one sublattice in one direction while
while freezing all other sublattice displacements.
Since the electric enthalpy
functional of Eq.~(\ref{eq:Ftotal}) is variational with respect to
the field-polarized Bloch functions under the constraints of
orthonormality, a constrained variational
principle exists for the second-order derivative of this functional
with respect to atomic displacements.\cite{gonze95} In particular,
the correct first-order perturbed wavefunctions $\psi^{(1)}_{m\k}$
can be obtained by minimizing the second-order expansion of the
total energy with respect to atomic displacements,
\begin{eqnarray}
\nonumber
F^{(2)}[\psi_{m\k}^{(0)};\E]&=&
   \min_{\psi^{(1)}}\bigg(E_{\rm KS}[\psi^{(0)}_{m\k};\psi^{(1)}_{m\k}]\\
&&-\Omega \Pmac[\psi^{(0)}_{m\k};\psi^{(1)}_{m\k}]\cdot \E\bigg)^{(2)},
\label{eq:Ftwo}
\end{eqnarray}
subject to the constraints
\begin{equation}
\langle \psi_{m\k}^{(0)}|\psi_{n\k}^{(1)}\rangle=0
\end{equation}
(where $m$ and $n$ run over occupied states). The fact that only
zero-order and first-order wavefunctions appear in Eq.~(\ref{eq:Ftwo})
is a consequence of the ``2$n$+1 theorem.''\cite{gonze95} 

Recalling that $|\psi_{n\k}^{(1)}\rangle$ is the first-order
wavefunction response to a small real
displacement $\lambda$ of basis atom ${\nu}$ along Cartesian
direction $\alpha$, we can expand the external potential as
\begin{equation}
v_{{\rm ext}}({\bf r} )=v_{{\rm ext}}^{(0)}({\bf r} )
+v^{(1)}_{{\rm ext},\nu \alpha}({\bf r})\lambda
+v^{(2)}_{{\rm ext},\nu \alpha}({\bf r})\lambda^{2}
+...
\end{equation}     
where 
\begin{equation}
v^{(1)}_{\rm ext,\nu \alpha}({\bf r})=\sum_{n}
    \frac{\partial v_{\rm ext}({\bf r})}{\partial R_{n\nu\alpha}}\;,
\end{equation}
\begin{equation}
v^{(2)}_{\rm ext,\nu \alpha}({\bf r})=\sum_{n}
    \frac{\partial^2 v_{\rm ext}({\bf r})}{\partial R_{n\nu\alpha}^2}\;,
\end{equation}
etc.  From this we shall construct the second-order energy $F^{(2)}$ of
Eq.~(\ref{eq:Ftwo}), which has to be minimized in order to find
$|\psi_{n\k}^{(1)}\rangle$.  The minimized value of $F^{(2)}$ gives,
as a byproduct, the diagonal element of the force-constant matrix
associated with displacement $\nu\alpha$.  Once the
$|\psi_{n\k}^{(1)}\rangle$ have been computed for all $\nu\alpha$, the
off-diagonal elements of the force-constant matrix can
be calculated using a version of the $2n+1$ theorem as will be described
in Sec.~\ref{sec:fcmatrix}.

\subsubsection{Discretized k mesh}
\label{sec:kmesh}

In practice, we always work on a discretized mesh of k-points, and we have to
take into account the orthogonality constraints among wavefunctions
at a given k-point on the mesh.
Here, we are following the ``perturbation expansion after discretization''
(PEAD) approach introduced in Ref.~\onlinecite{nunes01}.
That is, we write down the energy functional in its discretized form, and then
consistently derive perturbation theory from this energy functional. 
Introducing Lagrange multipliers $\Lambda_{\k,mn}$ to enforce the
orthonormality constraints
\begin{equation}
\langle \psi_{m\k}|\psi_{n\k}\rangle=\delta_{mn} ,
\end{equation}
where $\psi_{n\k}$ are the Bloch wavefunctions, and letting $N$ be the
number of k-points, the effective total-energy functional of
Eq.~(\ref{eq:Ftotal}) can be written as
\begin{equation}
F=F_{\rm KS}+F_{\rm BP}+F_{\rm LM}
\label{eq:Fbig}
\end{equation}
where $F_{\rm KS}=E_{\rm KS}$, $F_{\rm BP} =-\Omega \Pmac\cdot {\cal E}$,
and $F_{\rm LM}$ are the Kohn-Sham, Berry-phase, and Lagrange-multiplier
terms, respectively.  The first and last of these are given by
\begin{equation}
F_{\rm KS}=\frac{f}{N}\sum_{\k n}^{\rm occ}
   \langle \psi_{n\k}|T+v_{{\rm ext}}|\psi_{n\k}\rangle +E_{\rm Hxc}[n] ,
\label{eq:Fbiga}
\end{equation}
and
\begin{equation}
F_{\rm LM} = -\frac{f}{N}\sum_{\k,mn}^{\rm occ} \Lambda_{\k,mn}
(\langle \psi_{m\k}|\psi_{n\k}\rangle-\delta_{mn}) ,
\label{eq:Fbigc}
\end{equation}
where $N$ is the number of k-points in the BZ.  As for the Berry-phase
term, we modify the notation of Eq.~(\ref{eq:Pmac}) slightly to write
this as
\begin{equation}
F_{\rm BP}=-\frac{ef}{2\pi}\sum_{i=1}^{3}\frac{{\cal E} \cdot {\bf a}_i}
{N_{\perp }^{(i)}} \sum_{\k} D_{\k,\k+\g_i}
\label{eq:Fbigb}
\end{equation}
where
\begin{equation}
D_{\k\k'}=\textrm{Im}\ln\det S_{\k\k'} 
\label{eq:Ddef}
\end{equation}
and $\g_i$ is the reciprocal lattice mesh vector in lattice direction
$i$.  (That is, $\k$ and $\k+\g_i$ are neighboring k-points in one of
the $N_\perp^{(i)}$ strings of k-points running in the
reciprocal lattice direction conjugate to ${\bf a}_i$.) Recall that
the matrix of Bloch overlaps was defined in Eq.~(\ref{eq:Sdef}).

We now expand all quantities in orders of the perturbation, e.g.,
$\Lambda(\lambda)=\Lambda^{(0)}+\lambda \Lambda^{(1)}+\lambda^2
\Lambda^{(2)}+...$, etc.  Similarly, we expand
$S_{\k\k'}(\lambda)=S_{\k\k'}^{(0)}+\lambda S_{\k\k'}^{(1)}+
\lambda^2 S_{\k\k'}^{(2)}+...$, 
and we also define
\begin{equation}
Q^{\phantom{()}}_{\k'\k}=[S_{\k\k'}^{(0)}]^{-1}
\label{eq:Qdef}
\end{equation}
to be the inverse of the zero-order $S$ matrix.  Applying the
$2n+1$ theorem to Eq.~(\ref{eq:Fbig}), the variational second-order
derivative of the total-energy functional is
\begin{equation}
F^{(2)}=F^{(2)}_{\rm KS}+F^{(2)}_{\rm BP}+F^{(2)}_{\rm LM}
\label{eq:fparts}
\end{equation}
where
\begin{eqnarray}
F^{(2)}_{\rm KS}&=&\frac{1}{N}\sum_{{\bf k},m}^{\rm occ}
  \Big[\langle \psi_{m{\bf k}}^{(1)}|
  T^{(0)}+v_{{\rm ext}}^{(0)}|\psi_{m{\bf k}}^{(1)}\rangle
+\langle \psi_{m{\bf k}}^{(0)}|v_{{\rm ext}}^{(1)}|\psi_{m{\bf k}}^{(1)}\rangle
\nonumber\\
&& \hskip 1.0cm
+\langle \psi_{m{\bf k}}^{(1)}|v_{{\rm ext}}^{(1)}
    |\psi_{m{\bf k}}^{(0)}\rangle \Big]
   +E_{\rm Hxc}^{(2)}[n] \;,
\label{eq:ftwoa}
\\ \nonumber\\
F^{(2)}_{\rm BP}&=&- \frac{ef}{4\pi}\sum_{i=1}^{3}\frac{\E \cdot a_i }{N_{\perp }^{(i)}}
\sum_{\k}D^{(2)}_{\k,\k+\g_i} \;,
\label{eq:ftwob}
\\ \nonumber\\
F^{(2)}_{\rm LM}&=& \frac{1}{N}-\sum_{{\bf k},mn} \Big[ \,
   \Lambda^{(1)}_{{\bf k},mn} \, \big( \,
   \langle\psi_{m{\bf k}}^{(1)}|\psi_{n{\bf k}}^{(0)}\rangle
+\langle \psi_{m{\bf k}}^{(0)}|\psi_{n{\bf k}}^{(1)}\rangle \, \big)
\nonumber\\
&& \hskip 2.6cm
  +\Lambda^{(0)}_{{\bf k},mn}\langle
   \psi_{m{\bf k}}^{(1)}|\psi_{n{\bf k}}^{(1)}\rangle
 \, \Big]
\;.
\label{eq:ftwoc}
\end{eqnarray}
In the Berry-phase term, Eq.~(\ref{eq:ftwob}), we use the approach of
Ref.~\onlinecite{nunes01} to obtain the expansion of $\ln
\textrm{det}S_{\k\k'}$ with respect to the perturbation.  It then
follows that
\begin{equation}
D^{(2)}_{\k\k'} = \textrm{Im} \, {\rm Tr} \, [ \,
 2 S_{\k\k'}^{(2)}Q^{\phantom{()}}_{\k'\k}
  -S_{\k\k'}^{(1)}Q^{\phantom{()}}_{\k'\k}
       S_{\k\k'}^{(1)}Q^{\phantom{()}}_{\k'\k} \,]
\label{eq:dtwo}
\end{equation}
where $S^{(2)}$, $S^{(1)}$ and $Q$ are regarded as $L\times L$
matrices ($L$ is the number of occupied bands), matrix products
are implied, and Tr is a matrix trace running over the occupied bands.
Finally, in the Lagrange-multiplier term, Eq.~(\ref{eq:ftwoc}),
a contribution $ \Lambda^{(2)}_{{\bf k},mn}(\langle \psi_{m{\bf k}}^{(0)}|
\psi_{n{\bf k}}^{(0)}\rangle -\delta_{mn})$ has been dropped from
Eq.~(\ref{eq:ftwoc}) because the zero-order wavefunctions, which
have been calculated in advance, always satisfy the orthonormality
constraints $\langle \psi_{m{\bf k}}^{(0)}|\psi_{n{\bf k}}^{(0)}\rangle
 =\delta_{mn}$.  Moreover, the
zero-order Lagrange multipliers are made diagonal by a
rotation among zero-order wavefunctions at each k point, and the
first-order wavefunctions are made
orthogonal to the zero-order ones at each iterative step, so that
Eq.~(\ref{eq:ftwoc}) simplifies further to become just
\begin{equation}
F^{(2)}_{\rm LM} =-\epsilon_{m{\bf k}}\langle
   \psi_{m{\bf k}}^{(1)}|\psi_{m{\bf k}}^{(1)}\rangle \;.
\label{eq:ftwocp}
\end{equation}
Here, we have restored the notation $\epsilon_{m{\bf k}}=
\Lambda^{(0)}_{{\bf k},mm}$ for the diagonal zero-order Lagrange
multipliers.

\subsubsection{Conjugate-gradient minimization}
\label{sec:cg}

The second-order expansion of the electric enthalpy functional in
Eq.~(\ref{eq:fparts}) is minimized with respect to the first-order
wavefunctions using a ``band-by-band" conjugate-gradient
algorithm.\cite{dfpt1,payne92}
For a given point $\k$ and band $m$, the 
steepest-descent direction at iteration $j$ is 
$|\zeta_{m\k,j}\rangle=\partial F^{(2)}/\partial\langle u^{(1)}_{m{\bf k}}|$,
where $F^{(2)}$ is given by Eqs.~(\ref{eq:ftwoa}-\ref{eq:ftwob}) and
(\ref{eq:ftwocp}).  The derivatives of $F^{(2)}_{\rm KS}$ and
$F^{(2)}_{\rm LM}$ are straightforward; the new element in the presence
of an electric field is the term
\begin{equation}
\frac{\partial E_{\rm BP}^{(2)}}{\partial \langle u_{m\k}^{(1)}|}
= - \frac{ief}{4\pi}\sum_{i=1}^{3} \frac{{\cal E} \cdot
  {\bf a}_i}{N_{\perp }^{(i)}}   
\,\big(\,|{\cal D}_{m\k,\k+\g_i}\rangle-|{\cal D}_{m\k,\k-\g_i}\rangle\,\big)
\label{eq:gredient1}
\end{equation}
where
\begin{equation}
{\cal D}_{m\k\k'}= \left(\,
|u_{\k'}^{(1)}\rangle Q^{\phantom{()}}_{\k'\k}
- |u_{\k'}^{(0)}\rangle Q^{\phantom{()}}_{\k'\k}S_{\k\k'}^{(1)}
                        Q^{\phantom{()}}_{\k'\k} \,\right)_{m}
\;.
\label{eq:matmul}
\end{equation}
In this equation, $|u_{\k'}^{(1)}\rangle$ and $|u_{\k'}^{(0)}\rangle$
are regarded as vectors of length $L$ (e.g., $|u_{m\k'}^{(1)}\rangle$,
$m=1,L$), and vector-matrix and matrix-matrix products of dimension $L$
are implied inside the parentheses.  Also,
\begin{equation}
S_{\k\k',mn}^{(1)}=\langle u_{m\k}^{(0)}|u_{n\k'}^{(1)}\rangle
                  +\langle u_{m\k}^{(1)}|u_{n\k'}^{(0)}\rangle
\end{equation}
are the first-order perturbed overlap matrices at neighboring k-points.
The standard procedure translates the steepest-descent directions
$|\zeta_{m\k,j}\rangle$ into preconditioned conjugate-gradient search
directions $|\varphi_{m\k,j}\rangle$. An improved wavefunction
for iteration $j+1$ is then obtained by letting
\begin{equation}
|u^{(1)}_{m{\bf k},j+1}\rangle=|u^{(1)}_{m\k,j}\rangle
+\theta |\varphi_{m\k,j}\rangle \; ,
\end{equation}
where $\theta$ is a real number to be determined.
Since the $\theta$-dependence of $F^{(2)}(\theta)$ is  quadratic, the
minimum of $F^{(2)}$ along the conjugate-gradient direction is easily
determined to be
\begin{equation}
\theta_{\rm min}=-\frac{1}{2}\,\frac{dF^{(2)}}{d\theta}\bigg |_{\theta=0}
\, \bigg (\frac{d^2F^{(2)}}{d\theta^2}\bigg |_{\theta=0}\bigg)^{-1}\; .
\end{equation}

\subsubsection{Construction of the force-constant matrix}
\label{sec:fcmatrix}

To calculate phonon frequencies, we have to construct the force-constant 
matrix
\begin{equation}
\Phi_{\nu\alpha,\mu\beta}=\frac{\partial^2 F}{\partial R_{\nu\alpha}
\partial R_{\mu\beta}} \;.
\end{equation}
Each diagonal element $\Phi_{\mu\beta,\mu\beta}$ has already been obtained by
minimizing the $F^{(2)}$ in Eq.~(\ref{eq:fparts}) for the corresponding
perturbation $\mu\beta$.  The off-diagonal elements
$\Phi_{\nu\alpha,\mu\beta}$ can also be determined using only
the first-order wavefunctions $u^{(1)}_{m{\bf k},\mu\beta}$
using the (non-variational) expression
\begin{eqnarray}
\nonumber
\Phi_{\nu\alpha,\mu\beta} &=&\frac{2\Omega}{(2\pi)^3}\int_{\rm BZ}
\sum_{m}^{\rm occ}\bigg (\langle u_{m{\bf k}}^{(0)}|v_{{\rm ext},\nu\alpha}^{(1)}
+v_{{\rm Hxc},\nu\alpha}^{(1)}
| u^{(1)}_{m{\bf k},\mu\beta}\rangle \\
&&+\langle u_{m{\bf k}}^{(0)}|v_{{\rm ext},\nu\alpha,\mu\beta}^{(2)}|u_{m{\bf k }}^{(0)} \rangle 
\bigg )d\k+ \frac{1}{2}E_{\rm Hxc,\nu\alpha,\mu\beta}^{(2)}\
\label{eq:eod}
\end{eqnarray}
where $v_{{\rm ext},\nu\alpha}^{(1)} = \partial v_{\rm ext}/\partial
R_{\nu\alpha}$ etc.

\subsection{Nonzero wavevector case}
\label{sec:nonzeroq}

In the case of a phonon of arbitrary wavevector $\q$, the displacements
of the atoms are essentially of the form ${b}_{n \nu\alpha} = \lambda\
\exp(i\q\cdot {\bf t}_n)$, where $\lambda$ is a complex number.  However, 
a perturbation of this form does not lead by itself to a Hermitian 
perturbation of the Hamiltonian.  This is unacceptable, because we 
want the second-order energy to remain real,
so that it can be straightforwardly minimized.  Thus, we follow the
approach of Ref.~\onlinecite{dfpt1} and take the displacements to be
\begin{equation}
{ b}_{n \nu\alpha} = \lambda\,e^{i\q\cdot {\bf t}_n}
                + \lambda^*\,e^{-i\q\cdot {\bf t}_n}  \; ,
\end{equation}
leading to
\begin{eqnarray}
\nonumber
v_{{\rm ext}}({\bf r} )
&=&v_{{\rm ext}}^{(0)}({\bf r})+\lambda\,v^{(1)}_{{\rm ext},\nu \alpha,\q}({\bf r})
+ \lambda^{*}\,v^{(1)}_{{\rm ext},\nu \alpha,-\q}({\bf r})\\
\nonumber &+&
 \lambda^{2}\,v^{(2)}_{{\rm ext},\nu \alpha, \q, \q}({\bf r})
+{\lambda^{*}}^{2}\,v^{(2)}_{{\rm ext},\nu \alpha,-\q,-\q}({\bf r})\\
\nonumber &+&
 \lambda \lambda^{*}\,v^{(2)}_{{\rm ext},\nu \alpha, \q,-\q}({\bf r})
+\lambda^{*}\lambda\,v^{(2)}_{{\rm ext},\nu \alpha,-\q, \q}({\bf r})
\\
&+&...
\end{eqnarray}  
where
\begin{equation}
v^{(1)}_{{\rm ext},\nu\alpha,\pm \q}({\bf r})=\sum_n
    \frac{\partial v_{\rm ext}({\bf r})}{\partial R_{n\nu\alpha}}
    e^{\pm i\q\cdot{\bf t}_n} \; ,
\end{equation}
\begin{equation}
v^{(2)}_{{\rm ext},\nu\alpha,\pm \q,\pm \q}({\bf r})=\sum_{nm} \frac
  {\partial^2 v_{\rm ext}({\bf r})}{\partial R_{n\nu\alpha}\partial R_{m\nu\alpha}}
  \,e^{\pm i\q\cdot{\bf t}_n}\,e^{\pm i\q\cdot{\bf t}_m}\; ,
\end{equation}
etc.
Similarly, the field-dependent Bloch wavefunctions $\psi$ and enthalpy 
functional $F$ can also be expanded in terms of $\lambda$ and its 
hermitian conjugate as
\begin{equation}
\psi_{m\k}({\bf r})=\psi_{m\k}^{(0)}({\bf r})
   +\lambda\,\psi^{(1)}_{m\k,\q}({\bf r})
   +\lambda^*\,\psi^{(1)}_{m\k,-\q}({\bf r})+...
\label{eq:psione}
\end{equation}
and
\begin{eqnarray}
\nonumber
F[\E]&=&\lambda F^{(0)}[\E]+F_{\q}^{(1)}[\E]
+\lambda^*F_{{-\bf q}}^{(1)}[\E]\\
\nonumber &+&
\lambda^{2}F_{\q,\q}^{(2)}[\E]
+2\lambda \lambda^{*}F_{\q,{-\bf q}}^{(2)}[\E]\\
&+&{\lambda^{*}}^{2}F_{{-\bf q},{-\bf q}}^{(2)}[\E]
+... \;.
\label{eq:fqexpan}
\end{eqnarray}

The first-order wavefunctions in response to a perturbation
with wavevector $\q$ have translational properties
\begin{equation}
\psi^{(1)}_{m{\bf k},\q}({\bf r + \bf R})=e^{i(\bf k+\bf \q)\cdot\bf r}
\psi^{(1)}_{m{\bf k},\q}({\bf r})
\label{eq:kqbc}
\end{equation}
that differ from those of the zero-order wavefunctions
\begin{equation}
\psi^{(0)}_{m{\bf k}}({\bf r + \bf R})=
e^{i{\bf k}\cdot\bf r}\psi^{(0)}_{m{\bf k}}({\bf r})
\;.
\label{eq:kbc}
\end{equation}
As a result, we cannot simply work in terms of perturbed Bloch functions
or use the usual Berry-phase expression in terms of strings of
Bloch functions.
Also, in contrast to the $\q$=0 case, in which only one set of
first-order wavefunctions was needed, we now need to solve for
two sets $\psi^{(1)}_{m{\bf k},\pm \q}$ corresponding to the
non-Hermitian perturbation at wavevector $\q$ and its Hermitian
conjugate at wavevector $-\q$.\cite{dfpt1}

We now proceed to write out the second-order energy functional
$F^{(2)}[\psi^{(0)}_{m{\bf k}};\psi^{(1)}_{m{\bf k},\pm \q};\E]$,
corresponding to the sum of the quadratic terms in Eq.~(\ref{eq:fqexpan}), 
and minimize it simultaneously with respect to
$\psi^{(1)}_{m{\bf k},\q}$ and $\psi^{(1)}_{m{\bf k},-\q}$.

First, making the same decomposition as in Eq.~(\ref{eq:fparts}), we
find that the Kohn-Sham part is
\begin{equation}
F^{(2)}_{\rm KS}
= E_{\q,-\q}^{(2)}[\psi^{(0)}_{m{\bf k}};\psi^{(1)}_{{m{\bf k}},-\q}]
+ E_{-\q,\q}^{(2)}[\psi^{(0)}_{m{\bf k}};\psi^{(1)}_{{m{\bf k}},\q}]
\;,
\end{equation}
where
\begin{eqnarray}
\nonumber
E_{-\q,\q}^{(2)}
&=&\frac{2\Omega}{(2\pi)^3}\int_{\rm BZ}
\sum_{m}^{\rm occ}\bigg (\langle u_{m{\bf k},\q}^{(1)}|v_{{\rm ext},{\bf k}+\q,{\bf k}+\q}^{(0)}
| u^{(1)}_{m{\bf k},\q}\rangle\\
\nonumber
&&+\langle u_{m{\bf k},\q}^{(1)}|
v_{{\rm Hxc},{\bf k}+\q,{\bf k}+\q}^{(0)}
| u^{(1)}_{m{\bf k},\q}\rangle\\
\nonumber
&&+\langle u_{m{\bf k,\q}}^{(1)}|v_{{\rm ext},{\bf k+\q},{\bf k}}^{(1)}
+v_{{\rm Hxc},{\bf k+q},{\bf k}}^{(1)}
| u^{(0)}_{m{\bf k}}\rangle\\
\nonumber
&&+\langle u_{m{\bf k}}^{(0)}|v_{{\rm ext},{\bf k},{\bf k+q}}^{(1)}
+v_{{\rm Hxc},{\bf k},{\bf k+\q}}^{(1)}
| u^{(1)}_{m{\bf k},\q}\rangle \\
&&+\langle u_{m{\bf k}}^{(0)}|v_{{\rm ext},{\bf k},{\bf k}}^{(2)}|u_{m{\bf k }}^{(0)} \rangle \bigg)
d\k
+ \frac{1}{2}E_{\rm Hxc}^{(2)} \;.
\label{eq:nvef2}
\end{eqnarray}
Note that terms $E_{\q,\q}^{(2)}$ and $E_{-\q,-\q}^{(2)}$ vanish,
essentially because such terms transform like perturbations of
wavevector $\pm2\q$ which, except when $2\q$ equals a reciprocal
lattice vector, are inconsistent with crystal periodicity
and thus cannot appear in the energy expression.  (If $2\q$
is equal to a reciprocal lattice vector, $E_{\q,\q}^{(2)}$
and $E_{-\q,-\q}^{(2)}$ still vanish, as can be shown using
time-reversal symmetry.)

Second, we consider the Berry-phase coupling term.  The treatment of this
term is rather subtle because, as mentioned above, the perturbed
wavefunctions are now admixtures of parts with periodicity as in
Eq.~(\ref{eq:kqbc}) and as in Eq.~(\ref{eq:kbc}), so that the
usual Berry-phase formula for the polarization\cite{smith93}
cannot be used.  A different approach is needed now in order to
express the polarization in terms of the perturbed wavefunctions.
For this purpose, we consider a virtual supercell in which the
wavevectors $\k$ and $\q$ would be commensurate, and make
use of the definition introduced by Resta\cite{resta98} specialized
to the non-interacting case.  The details of this treatment are
deferred to the Appendix, but the results can be written in the
relatively simple form
\begin{equation}
F^{(2)}_{\rm BP}=- \frac{ef}{2\pi}\sum_{i=1}^{3}\frac{\E \cdot a_i }{N_{\perp }^{(i)}}
\sum_{\k}D^{(2)}_\k(\g_i)
\label{eq:f2}
\end{equation}
where
\begin{eqnarray}
&&D^{(2)}_\k(\g)= \textrm{Tr} \, \big[ \,
   S^{(1,1)}_{\k,\k+\g} Q^{\phantom{()}}_{\k+\g,\k}
 - S^{(1,0)}_{\k,\k+\g-\q} 
\nonumber\\
 &&\qquad  
 \times \,                 Q^{\phantom{()}}_{\k+\g-\q,\k-\q}
   S^{(0,1)}_{\k-\q,\k+\g} Q^{\phantom{()}}_{\k+\g,\k} \, \big]
\label{eq:loops}
\end{eqnarray}
with $Q_{\k'\k}$ given by Eq.~(\ref{eq:Qdef}) and the superscript
notation
$S^{(s,t)}={\partial^{s+t} S}/{\partial({\lambda^{*}})^s \partial 
{\lambda^t}}$.  From Eqs.~(\ref{eq:Sdef}) and (\ref{eq:kqbc}), we can
write these explicitly as
\begin{eqnarray}
S^{(1,0)}_{\k\k',mn} &=&
  \langle\psi_{m\k}^{(0)}|e^{-i\g\cdot{\bf r}}|\psi_{n\k',\q}^{(1)}\rangle
\nonumber\\ &&\quad
 +\langle\psi_{m\k,-\q}^{(1)}|e^{-i\g\cdot{\bf r}}|\psi_{n\k'}^{(0)}\rangle
\;,
\end{eqnarray}
\begin{eqnarray}
S^{(0,1)}_{\k\k',mn} &=&
  \langle\psi_{m\k}^{(0)}|e^{-i\g\cdot{\bf r}}|\psi_{n\k',-\q}^{(1)}\rangle
\nonumber\\ &&\quad
 +\langle\psi_{m\k,\q}^{(1)}|e^{-i\g\cdot{\bf r}}|\psi_{n\k'}^{(0)}\rangle
\;,
\end{eqnarray}
\begin{eqnarray}
S^{(1,1)}_{\k\k',mn} &=&
  \langle\psi_{m\k,\q}^{(1)}|e^{-i\g\cdot{\bf r}}|\psi_{n\k',\q}^{(1)}\rangle
\nonumber\\ &&\quad
 +\langle\psi_{m\k,-\q}^{(1)}|e^{-i\g\cdot{\bf r}}|\psi_{n\k',-\q}^{(1)}\rangle
\;.
\end{eqnarray}

Third, the treatment of the Lagrange-multiplier term is straightforward;
in analogy with Eq.~(\ref{eq:ftwocp}), we obtain
\begin{equation}
F^{(2)}_{\rm LM} = -\epsilon_{m{\bf k}}\big (\langle
   \psi_{m{\bf k},\q}^{(1)}|\psi_{m{\bf k},\q}^{(1)}\rangle 
   +\langle
   \psi_{m{\bf k},-\q}^{(1)}|\psi_{m{\bf k},-\q}^{(1)}\rangle \big )\;.
\label{eq:ftwocp2}
\end{equation}

\begin{figure}
\begin{center}
   \epsfig{file=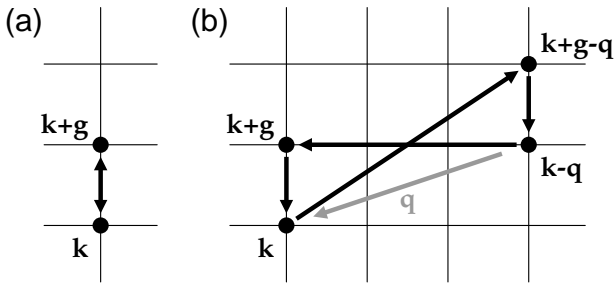,width=3.2in}
\end{center}
\caption{Pattern of couplings between k-points arising
in (a) the first term, and (b) the second term, of
Eq.~(\protect\ref{eq:loops}). Reciprocal vector $\bf q$ is the
phonon wavevector, while $\bf g$ is a primitive vector of the
k-point mesh (indicated by thin horizontal and vertical lines).}
\label{fig:kpoint}
\end{figure}

If we look closely at Eq.~(\ref{eq:loops}), we see that the second
term involves not simply pairs of k-points separated by the mesh
vector $\g$, but {\it quartets} of k-points, as illustrated in
Fig.~\ref{fig:kpoint}.  Reading from left to right in the second
term of Eq.~(\ref{eq:loops}), the k-point labels are $\k$, then
$\k+\g-\q$, then $\k-\q$, then $\k+\g$, and finally back to $\k$.
This is the loop illustrated in Fig.~\ref{fig:kpoint}.  Each dark arrow
represents a matrix element of $S^{(1,0)}$, $S^{(0,1)}$, or $Q$;
the gray arrow indicates the phonon $\q$-vector.
These loops arise because there are two kinds of coupling between
k-points entering into the present theory.  First, even in the
absence of the phonon perturbation, wavevectors at neighboring
k-points separated by mesh vector $\g$ are coupled by the
$\E\cdot{\bf P}$ term in the energy functional.  Second, the
phonon introduces a perturbation at wavevector $\q$.  It is the
interplay between these two types of inter-k-point coupling
that is responsible for the appearance of these four-point loops
in the expression for $F^{(2)}_{\rm BP}$.

The implementation of the conjugate-gradient minimization
algorithm proceeds in a manner very similar to that outlined in
Sec.~\ref{sec:cg}.
Naively, one would have to work simultaneously with the two
search-direction vectors
\begin{eqnarray}
|\zeta_{m\k,\q}\rangle &=& \partial F^{(2)}/\partial\langle u^{(1)}_{m\k,\q}| \;,
\nonumber\\
|\zeta_{m\k,-\q}\rangle &=& \partial F^{(2)}/\partial\langle u^{(1)}_{m\k,-\q}| \;,
\label{eq:sdv}
\end{eqnarray}
where $u^{(1)}_{m\k,\pm\q}$ are the periodic parts of $\psi_{m\k,\pm\q}^{(1)}$.
However, minimizing the second-order energy $F^{(2)}$ with respect
to two sets of first-order wavefunctions $u_{n\k,\pm \q}$
would double the computational cost and would involve substantial
restructuring of existing computer codes.  We can avoid this by using
the fact that the second-order energy is invariant under time reversal
to eliminate one set of first-order wavefunctions
$\psi_{n\k,-\q}^{(1)}$ in favor of the other set $\psi_{n\k,\q}^{(1)}$
following the approach given in Ref.~\onlinecite{dfpt1}.
Specifically, the two sets of first-order wavefunctions are related by
\begin{eqnarray}
\psi_{n\k}^{(0)}(\bf r)&=&e^{i\theta_{n\k}}\psi_{n-\k}^{(0)*}(\bf r)\;,\\
\psi_{n\k, \q}^{(1)}\bf r)&=&e^{i\theta_{n\k}}\psi_{n-\k,- \q}^{(1)*}(\bf r)\;,
\end{eqnarray}
where $\theta_{n\k}$ is an arbitrary phase independent of
${\bf r}$. The arbitrary phase $\theta_{n\k}$ cancels out in
the expression of $F^{(2)}$ since every term in $F^{(2)}$ is
independent of the phase of the first-order wavefunctions. Thus, we
choose $\theta_{n\k}=1$ for simplicity and write the second-order
energy functional in terms of wavefunctions $\psi_{n\k,\q}$ only.

The minimization procedure now proceeds in a manner similar to the
zero-wavevector case, except that
the calculation of the Berry-phase part involves some vector-matrix-matrix
products as in Eq.~(\ref{eq:matmul}), but circulating around three
of the sides of the loop in Fig.~\ref{fig:kpoint}.
Since $F^{(2)}$ remains in a quadratic form, the minimum of $F^{(2)}$ is
again easily searched along the conjugate-gradient direction.  Wavefunctions
are updated over k-points one after another, and the first-order
wavefunctions are updated. This
procedure continues until the self-consistent potential is
converged. Once the first-order responses of wavefunctions are
obtained, the diagonal elements of the dynamical matrix are obtained
by evaluating $F^{(2)}$, and the off-diagonal elements are obtained
from a straightforward generalization of Eq.~(\ref{eq:eod}),
\begin{eqnarray}
\nonumber
\Phi_{\nu\alpha,\mu\beta} &=&\frac{2\Omega}{(2\pi)^3}\int_{\rm BZ}
\sum_{m}^{\rm occ}\bigg(\langle u_{m{\bf k}}^{(0)}|v_{{\rm ext},\nu\alpha,\k,\k+\q}^{(1)}
| u^{(1)}_{m{\bf k},\mu\beta,\q}\rangle\\
\nonumber
&&+\langle u_{m{\bf k}}^{(0)}|
v_{{\rm Hxc},\nu\alpha,\k,\k+\q}^{(1)}
| u^{(1)}_{m{\bf k},\mu\beta,\q}\rangle \\
&&+\langle u_{m{\bf k}}^{(0)}|v_{{\rm ext},\nu\alpha,\mu\beta}^{(2)}|u_{m{\bf k }}^{(0)} \rangle 
\bigg )d\k
\nonumber\\
&&+ \frac{1}{2}{E^{(2)}_{\rm Hxc,\nu\alpha,\mu\beta}}\;.
\label{eq:eod2}
\end{eqnarray}

\section{Test calculations for III-V semiconductors}
\label{sec:test}

In order to test our method, we have carried out calculations of
the frequency shifts induced by electric fields in two III-V
semiconductors, AlAs and GaAs.  We have chosen
these two materials because they are well-studied systems both
experimentally and theoretically, and because the symmetry allows
some phonon mode frequencies to shift linearly with electric
field while others shift quadratically.  Since our main purpose
is to check the internal consistency of our theoretical approach,
we focus on making comparisons between the shifts calculated using
our new linear-response method and those calculated using standard
finite-difference methods.  Moreover, as mentioned at the start of
Sec.~\ref{sec:exact}, we have chosen to neglect changes in phonon
frequencies that enter through the electric-field induced strains
(piezoelectric and electrostrictive effects), and we do this
consistently in both the linear-response and finite-difference
calculations.  For this reason, our results are not immediately
suitable for comparison with experimental measurements.

Our calculations are carried out using a plane-wave pseudopotential
approach to density-functional theory.  We use the ABINIT
code package,\cite{abinit} which incorporates the finite
electric field method of Souza {\it et al.}\cite{souza02}
for the ground-state and frozen-phonon calculations in finite electric
field.  We then carried out the linear-response calculations with a
version of the code that we have modified to implement
the linear-response formulas of the previous section.

The details of the calculations are as follows.  We use
Troullier-Martins norm-conserving pseu\-do\-po\-tentials,\cite{troullier91}
the Teter Pade parameterization\cite{teter96} of the
local-density approximation, and a plane-wave cutoff of 16 Hartree.
A 10$\times$10$\times$10 Monkhorst-Pack\cite{monkhorst76} k-point sampling
was used, and we chose lattice constants of 10.62\,\AA\ and 10.30\,\AA\ for
AlAs and GaAs, respectively.
The crystals are oriented so that the vector $(a/2)(1,1,1)$ points from 
a Ga or Al atom to an As atom.

\begin{table}
\caption{Calculated frequency shifts, in cm$^{-1}$, induced by an
electric field of $5.14 \times 10^8$\,V/m applied along $x$
in GaAs and AlAs.  `FD' are the results of finite-difference
(frozen-phonon) calculations in which atoms are displaced
by hand and restoring forces are calculated, while `LR' refers to 
the use of the linear-response developed here.  The L and X points
are at $(2\pi/a)(1,1,1)$ and $(2\pi/a)(1,0,0)$ respectively.}
\begin{ruledtabular}
\begin{tabular}{lrrrr}
 & \multicolumn{2}{c}{GaAs} & \multicolumn{2}{c}{AlAs} \cr
 Mode & FD\;\; & LR\;\; & FD\;\; & LR\;\; \cr
\hline
 $\Gamma$ O1 \footnotemark[1]  & $-$3.856& $-$3.856 & $-$5.941 & $-$5.941   \cr
 $\Gamma$ O2 \footnotemark[1]  & $-$0.282& $-$0.281 & $-$0.300& $-$0.299 \cr
 $\Gamma$ O3 \footnotemark[1]  & 3.548 & 3.548 & 5.647 & 5.647 \cr
 \hline
 L LO & 2.701 & 2.703& 4.282&4.282 \cr
 L TO1 & $-$3.749&$-$3.749& $-$5.663&$-$5.663 \cr
 L TO2 & 0.567&0.564& 0.952 & 0.952 \cr
 \hline
 X LO & 0.050 & 0.050 & $-$0.243 & $-$0.243 \cr
 X TO1 & $-$3.953& $-$3.953 & $-$6.083 & $-$6.083 \cr
 X TO2 & 3.753 & 3.753& 5.919 &5.919 \cr
\end{tabular}
\end{ruledtabular}
\label{table:1}
\footnotetext[1]{The non-analytic long-range Coulomb contributions are
excluded for the $\Gamma$ modes.}
\end{table}

Table \ref{table:1} shows the changes in phonon frequencies resulting from
an electric field applied along a Cartesian direction at several
high-symmetry q-points in GaAs and AlAs.  Both the electronic and
ionic contributions, Eqs.~(\ref{eq:wel}-\ref{eq:wion}), are included.
We first relaxed the atomic
coordinates in the finite electric field until the maximum force
on any atom was less than $10^{-6}$\,Hartree/Bohr.
We then carried out the linear-response calculation, and in addition,
to check the internal consistency of our linear-response method,
we carried out a corresponding calculation using a finite-difference
frozen-phonon approach.  For the latter, the atoms were displaced according
to the normal modes obtained from our linear-response calculation,
with the largest displacement being 0.0025 Bohr.
(Because the electric field lowers the symmetry, the symmetry-reduced
set of k-points is not the same as in the absence of the electric
field.)  The agreement between the finite-different approach and the
new linear-response implementation can be seen to be excellent, with
the small differences visible for some modes being attributable to
truncation in the finite-difference formula and the finite density
of the k-point mesh.

In Table \ref{table:twopart}, we decompose the frequency shifts into
the ionic contribution $\Delta\o_{\rm ion}(\q;\E)$
and the electronic contribution $\Delta\o_{\rm el}(\q;\E)$
defined by Eqs.~(\ref{eq:wion}) and (\ref{eq:wel}), respectively,
calculated using the linear-response approach.  It is clear that the
largest contributions are ionic in origin.  For example, the large,
roughly equal and opposite shifts of the O1 and O3 modes at $\Gamma$
arise from the ionic terms.  However, there are special cases (e.g., O2 at
$\Gamma$ and LO at X) for which the ionic contribution happens to be
small, so that the electronic contribution is comparable in magnitude.

The pattern of ionic splittings appearing at $\Gamma$ can be
understood as follows.  Because the non-analytic long-range Coulomb
contribution is not included, the three optical modes at $\Gamma$ are initially
degenerate with frequency $\omega_0$ in the unperturbed lattice.  A
first-order electric field along $x$ induces a first-order relative
displacement $u_x$ of the two sublattices, also along $x$.  By symmetry
considerations, the perturbed dynamical matrix is given, up to quadratic
order in $u_x$, as
\begin{equation}
D(\Gamma)=\omega_0^2\left ( 
\begin{array}{ccc}
1+\mu u_x^2 & 0 & 0  \\
0 & 1+\nu u_x^2 & \kappa u_x \\
0& \kappa u_x & 1+\nu u_x^2 \\
\end{array}
\right ) \;.
\end{equation}
The off-diagonal $\kappa$ term arises from the $E_{xyz}$ coupling
in the expansion of
the total energy in displacements; this is the only third-order term
allowed by symmetry.  The $\mu$ and $\nu$ terms arise from fourth-order
couplings of the form $E_{xxxx}$ and $E_{xxyy}$ respectively.
The eigenvalues of this matrix are proportional to $1+\mu u_x^2$ and
$1\pm\kappa u_x+\nu u_x^2$.  Thus, two of the modes should be
perturbed at first order in the field-induced displacements
with a pattern of equal and opposite frequency shifts, while all three
modes should have smaller shifts arising from the quadratic terms.
This is just what is observed in the pattern of frequency shifts shown
in Table \ref{table:twopart}.  (The symmetry of the pattern of electronic
splittings is the same, but it turns out that the linear shift is
much smaller in this case, so that for the chosen electric field,
the linear and quadratic contributions to the electronic frequency
shift have similar magnitudes.) A similar analysis can be used to
understand the patterns of frequency shifts at the $L$ and $X$ points.

\begin{table}
\caption{Same as in Table \ref{table:1}, but with the
frequency shifts decomposed into ionic and electronic contributions
as defined in Eqs.~(\ref{eq:wion}) and (\ref{eq:wel}) respectively.}
\begin{ruledtabular}
\begin{tabular}{lrrrr}
 & \multicolumn{2}{c}{GaAs} & \multicolumn{2}{c}{AlAs} \cr
& Ion \; & Elec. & Ion \;& Elec.  \cr
\hline
$\Gamma$ O1 \footnotemark[1] & $-$3.659&$-$0.198  &$-$5.684 & $-$0.257  \cr
$\Gamma$ O2 \footnotemark[1] & $-$0.146& $-$0.135& $-$0.123 & $-$0.177\cr
$\Gamma$ O3 \footnotemark[1] & 3.655 & $-$0.107& 5.589 & 0.058 \cr
\hline
L LO &  2.341 & 0.362 &3.633 & 0.649 \cr
L TO1 & $-$3.486 &	$-$0.262  & $-$5.628 & $-$0.034\cr
L TO2 &  1.181 &	$-$0.617 &  1.658 & $-$0.707\cr
 \hline
 X LO & 0.122	& $-$0.073 & $-$0.033 &	$-$0.209 \cr
 X TO1 & $-$3.411	& $-$0.543 & $-$5.658 &	$-$0.424 \cr
 X TO2 & 3.388 &	0.365 &  5.609	& 0.310\cr
\end{tabular}
\end{ruledtabular}
\label{table:twopart}
\footnotetext[1]{The non-analytic long-range Coulomb contributions are
excluded for the $\Gamma$ modes.}
\end{table}

\begin{figure}
\begin{center}
   \epsfig{file=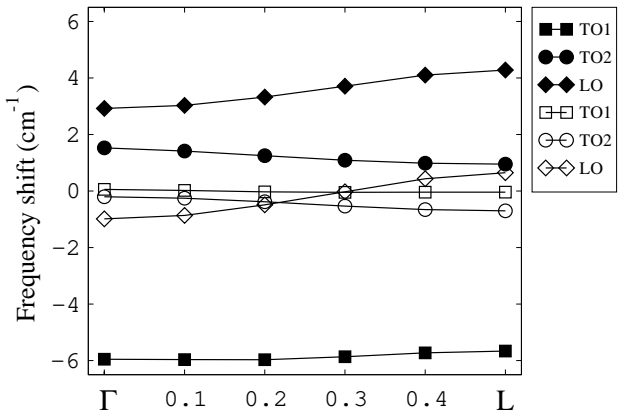,width=3.2in}
\end{center}
\caption{Frequency shifts induced by an electric field of
5.14$\times 10^8$\,V/m along $x$ in AlAs, plotted along $\Gamma$ to $L$.
Filled and open symbols indicate the total shift
$\Delta\o_{\rm el}+\Delta\o_{\rm ion}$ and the electronic contribution
$\Delta\o_{\rm el}$ respectively.}
\label{fig:test}
\end{figure}

We have also plotted, in Fig.~\ref{fig:test}, the calculated total
frequency shift $\Delta\o_{\rm el}(\q)+\Delta\o_{\rm ion}(\q)$
and its electronic contribution $\Delta\o_{\rm el}(\q)$ along
the line from $\Gamma$ to L for the case of AlAs.  (The `LO' and
`TO' symmetry labels are not strictly appropriate here because the
electric field along $x$ mixes the mode eigenvectors; the notation
indicates the mode that would be arrived at by turning off the field.)
In contrast to the results presented in Tables
\ref{table:1}-\ref{table:twopart}, the frequencies at $\Gamma$
in Fig.~\ref{fig:test} were computed by
including the long-range non-analytic Coulomb contribution for
$\hat{q}\parallel(111)$ in order to extend the curves to $q=0$.
(Because the direct linear-response calculation of the dynamical
effective charge and dielectric susceptibility tensors are not yet
developed and implemented in the presence of a finite electric
field, the needed tensor elements were computed by finite differences.)
It is clearly evident that the electronic terms remain much smaller than the
ionic ones for all three optical modes over the entire branch in
$q$-space.

\begin{table}[b]
\caption{Comparison of ionic displacements and frequency shifts at the
$L$ point in GaAs as computed by the approximate and exact approaches of
Sec.~\ref{sec:approx} and \ref{sec:exact} respectively, again for an
electric field of $5.14 \times 10^8$\,V/m along $x$.  $R_{\E}$ is the
induced displacement of the cation sublattice along $x$, and the
$\Delta\o_{\rm ion}$ are ionic contributions to the frequency shifts as
defined in Eq.~(\ref{eq:wion}).}
\begin{ruledtabular}
\begin{tabular}{llcrrr}
& & \multicolumn{1}{c}{$R_{\E}$} &
      \multicolumn{3}{c}{$\Delta\o_{\rm ion}$(L) (cm$^{-1}$)}\cr
& & (10$^{-3}$\,\AA) & LO & TO1 & TO2 \cr
 \hline
GaAs & Approx. & 5.07 & 2.63  & $-$3.89 & 1.37 \cr
     & Exact   & 4.95 & 2.34  & $-$3.49 & 1.18 \cr
AlAs & Approx. & 5.69 & 3.75  & $-$5.66 & 1.65 \cr
     & Exact   & 5.62 & 3.63 & $-$5.63  & 1.66  \cr
\end{tabular}
\end{ruledtabular}
\label{table:three}
\end{table}

Returning now to the comparison between our exact theory of
Sec.~\ref{sec:exact} and the approximate theory of Sec.~\ref{sec:approx},
we compare the equilibrium positions and phonon frequencies predicted
by these theories in Table \ref{table:three}.  Recall that $R_{\E}$
is calculated in the approximate theory by using Eq.~(\ref{eq:forcebal}).
Using this force, the ion coordinates were again relaxed to a
tolerance of $10^{-6}$ (Hartree/Bohr) on the forces.  It can be seen
that $R_{\E}$ is predicted quite well by the approximate theory, with
errors of only $\sim$2\%, confirming that the displacements can be
calculated to good accuracy using a linearized theory for this magnitude
of electric field.  The changes in the phonon frequencies resulting from
these displacements (evaluated at zero and non-zero field for the
approximate and exact theories respectively) are listed in
the remaining columns of Table \ref{table:three}.  The discrepancies
in the phonon frequencies are now somewhat larger, approaching 15\% in
some cases.  This indicates that the approximate theory is able to give
a moderately good description of the phonon frequency shifts of
GaAs in this field range, but the exact theory is needed for
accurate predictions.  (Also, recall that the approximate theory does
not provide any estimate for the electronic contributions, which are
not included in Table \ref{table:three}.)

\begin{figure}
\begin{center}
   \epsfig{file=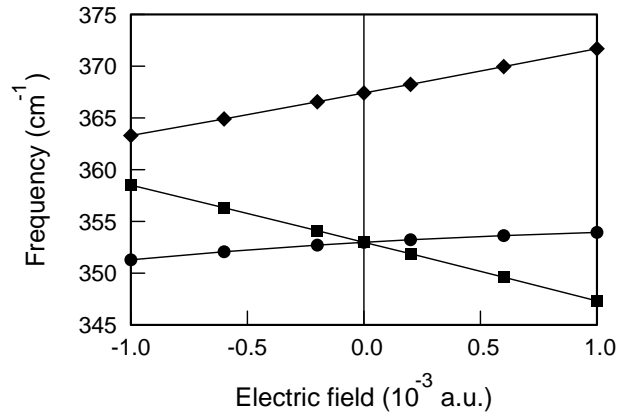,width=3.2in}
\end{center}
\caption{Frequencies of LO and TO modes at $L$ in AlAs as a
function of electric field (where 10$^{-3}$\,a.u.~=~$5.14 \times
10^8$\,V/m) applied along $x$.  The symbols have the same
interpretation as in Fig.~\ref{fig:test}.}
\label{fig:nonlin}
\end{figure}

Finally, we illustrate our ability to calculate the nonlinear field
dependence of the phonon frequencies by presenting the calculated optical
$L$-point phonon frequencies of AlAs in Fig.~\ref{fig:nonlin} as a
function of electric field along $x$.  These are again the results of
our exact theory, obtained by including both ionic and electronic
contributions.  The two TO modes are degenerate at zero field, as
they should be.  All three modes show a linear component that
dominates their behavior in this range of fields.  However, a quadratic
component is also clearly evident, illustrating the ability of the
present approach to describe such nonlinear behavior.

\section{Summary and discussion}
\label{sec:summa}

We have developed a method for computing the phonon frequencies of
an insulator in the presence of a homogeneous, static electric field.
The extension of density-functional perturbation theory to this case
has been accomplished by carrying out a careful expansion of the
field-dependent energy functional $E_{\rm KS}+\Omega\E \cdot {\bf P }$, where
${\bf P}$ is the Berry-phase polarization, with respect to phonon
modes both at $q=0$ and at arbitrary $q$.  In the general case of
nonzero $q$, there is a subtle interplay between the couplings between
neighboring k-points introduced by the electric field and the
further-neighbor couplings introduced by the $q$-vector, so that
terms arise that require the evaluation of four-sided loops in k-space.
However, with the judicious use of time-reversal symmetry, the
needed evaluations can be reduced to a form that is not difficult to
implement in an existing DFPT code.

We have carried out test calculations on two III-V semiconductors, AlAs
and GaAs, in order to test the correctness of our implementation.
A comparison of the results of linear-response and finite-difference
calculations shows excellent agreement, thus validating our approach.
We also decompose the frequency shifts into ``lattice'' and
``electronic'' contributions and quantify these, and we find that the
lattice contributions (i.e., those resulting from induced displacements
in the reference equilibrium structure) are usually, but not always,
dominant.  We also evaluated the accuracy of an approximate method
for computing the lattice contribution, in which only zero-field
inputs are needed.  We found that this approximate approach gives
a good rough description, but that the full method is needed for an
accurate calculation.

Our linear-response method has the same advantages, relative to the
finite-difference approach, as in zero electric field.  Even
for a phonon at $\Gamma$, our approach is more direct and simplifies
the calculation of the phonon frequencies.  However, its real advantage
is realized for phonons at arbitrary $q$, because the frequency
can still be obtained efficiently from a calculation on a single unit
cell without the need for imposing commensurability of the $q$-vector
and computing the mode frequencies for the corresponding supercell.
We also emphasize that the method is not limited to infinitesimal
electric fields.  We thus expect the method will prove broadly
useful for the study of linear and nonlinear effects of electric
bias on the lattice vibrational properties of insulating materials.

\acknowledgments

This work was supported by NSF grants DMR-0233925 and DMR-0549198. 
We wish to thank I.~Souza for assistance in the early stages of the
project.

\appendix
\section*{Appendix}
\label{app:appen}

The formula for the electric polarization given in the original work
of King-Smith and Vanderbilt\cite{smith93} is not a suitable starting
point for the phonon perturbation analysis that we wish to derive here,
because a perturbation of nonzero wavevector $\q$ acting on a Bloch
function generates a wavefunction that is no longer of Bloch form.
That is, while the zero-order wavefunction $\psi^{(0)}_{m\k}$ transforms
as $e^{i\k\cdot\R}$ under a translation by $\R$, the first-order
wavefunction $\psi^{(1)}_{m\k,\q}$ transforms as $e^{i(\k+\q)\cdot\R}$.

To solve this problem, we first restrict ourselves to the case of a
regular mesh of $N=N_1\times N_2\times N_3$ k-points in the Brillouin
zone.  As is well known, one can regard the Bloch functions at these
k-points as being the solutions at a single k-point of the downfolded
Brillouin zone of an  $N_1\times N_2\times N_3$ supercell.  Then, as
long as the wavevector $\q$ is a reciprocal lattice vector of the
supercell, or $\q=m_1{\bf b}_1/N_1+m_2{\bf b}_2/N_2+m_3{\bf b}_3/N_3$,
the phonon perturbation will be commensurate with
the supercell, and the perturbed wavefunction will continue to be a
zone-center Bloch function of the supercell.  We thus restrict our
analysis to this case.

A formula for the Berry-phase polarization for single-k-point
sampling of a supercell has been provided by Resta.\cite{resta98}
Starting from a general many-body formulation in terms of a
definition of the position operator suitable for periodic boundary
conditions, and then specializing to the case of a single-particle
Hamiltonian, Resta's derivation leads to
\begin{equation}
{\bf P}= \frac{ef}{2\pi\Omega} \sum_\alpha \gamma_\alpha \, {\bf a}_\alpha
\end{equation}
where the Berry phase in lattice direction $\alpha$ is given by
\begin{equation}
\gamma_\alpha=-\textrm{Im}\, \ln \det M_\alpha \;.
\end{equation}
Here
\begin{equation}
M_{\alpha,ss'}=\langle \psi_s|e^{-i{\bf g}_\alpha\cdot\rr} |\psi_{s'} \rangle
\label{eq:MM}
\end{equation}
where ${\bf g}_\alpha={\bf b}_\alpha/N_\alpha$ is the primitive reciprocal
mesh vector in lattice direction $\alpha$
and $s$ runs over all of the occupied states of the supercell.
Expanding the matrix $M_\alpha$ in powers of $\lambda$
and $\lambda^*$,
\begin{eqnarray}
M_\alpha(\lambda,\lambda^*) &=&
M_\alpha^{(0,0)}
+\lambda M_\alpha^{(1,0)}
+\lambda^* M_\alpha^{(0,1)}
+\lambda^2 M_\alpha^{(2,0)}
\nonumber\\
&&+|\lambda|^2 M_\alpha^{(1,1)}
+\lambda^{*2} M_\alpha^{(0,2)}
+... \;,
\end{eqnarray}
the expansion of $\ln \det M_\alpha(\lambda,\lambda^*)$ takes the form
\cite{nunes01}
\begin{eqnarray}
\ln \det M_\alpha&=&\ln \det M_\alpha^{(0,0)}\nonumber\\
&+&\lambda\,\textrm{Tr}[M_\alpha^{(1,0)}Q_\alpha]
+\lambda^*\,\textrm{Tr}[M_\alpha^{(0,1)}Q_\alpha]\nonumber\\
&+&\lambda^2\,\textrm{Tr}[2M_\alpha^{(2,0)}Q_\alpha
-M_\alpha^{(1,0)}Q_\alpha M_\alpha^{(1,0)}Q_\alpha]\nonumber\\
&+&\lambda^{*2}\,\textrm{Tr}[2M_\alpha^{(0,2)}Q_\alpha
-M_\alpha^{(0,1)}Q_\alpha M_\alpha^{(0,1)}Q_\alpha]\nonumber\\
&+&|\lambda|^2\,\textrm{Tr}[2M_\alpha^{(1,1)}Q_\alpha
-M_\alpha^{(1,0)}Q_\alpha M_\alpha^{(0,1)}Q_\alpha\nonumber\\
&&\qquad\qquad-M_\alpha^{(0,1)}Q_\alpha M_\alpha^{(1,0)}Q_\alpha]\nonumber\\
&+& \textrm{higer order terms} \;,
\end{eqnarray}
where $Q_{\alpha}=[M_{\alpha}^{(0,0)}]^{-1}$.

From the physical point of view, the terms proportional to $\lambda$,
$\lambda^*$, $\lambda^{2}$, and $\lambda^{*2}$ should vanish as a result of
translational symmetry.  For example, a term linear in $\lambda$ should
transform like $e^{i\q\cdot\R}$ under translation by a lattice vector
$\R$, but such a form is inappropriate in an expression for the energy,
which must be an invariant under translation.  We have confirmed this
by explicitly carrying out the matrix multiplications for these terms
and checking that the traces are zero.  Using the cyclic property
of the trace to combine the last two terms, we find that the
overall second-order change in $\ln \det M_\alpha$ is
\begin{eqnarray}
(\ln \det M_\alpha)^{(2)}&=& 2|\lambda|^2\,\textrm{Tr}[M_\alpha^{(1,1)}Q_\alpha
\nonumber\\
&&\quad-M_\alpha^{(1,0)}Q_\alpha M_\alpha^{(0,1)}Q_\alpha]
\;.
\label{eq:logdeta}
\end{eqnarray}

In our case, the orbitals $\psi_s$ appearing in Eq.~(\ref{eq:MM}) are the
perturbed wavefunctions originating from the unperturbed states
labeled by band $m$ and k-point $\k$ of the primitive cell, so
that we can let $\psi_s\rightarrow\psi_{m\k}$ and
\begin{equation}
M_{\alpha,m\k,m'\k'}=\langle \psi_{m\k}|e^{-i{\bf g}_\alpha\cdot\rr}
|\psi_{m'\k'} \rangle \;.
\label{eq:Mexpan}
\end{equation}
Substituting Eq.~(\ref{eq:psione}) into Eq.~(\ref{eq:Mexpan}), we find
\begin{eqnarray}
M_{\alpha,m\k,m'\k'}^{(1,0)} &=&
\langle \psi_{m\k}^{(0)}|e^{-i{\bf g}_\alpha \cdot {\bf r}} |\psi_{m'\k',\q}^{(1)} \rangle
\nonumber\\ &&+
\langle \psi_{m\k,-\q}^{(1)}|e^{-i{\bf g}_\alpha \cdot {\bf r}}|\psi_{m'\k'}^{(0)} \rangle \;,
\label{eq:zba}
\end{eqnarray}
\begin{eqnarray}
M_{\alpha,m\k,m'\k'}^{(0,1)} &=&
\langle \psi_{m\k}^{(0)}|e^{-i{\bf g}_\alpha \cdot {\bf r}} |\psi_{m'\k',-\q}^{(1)}\rangle
\nonumber\\ &&+
\langle \psi_{m\k,\q}^{(1)}|e^{-i{\bf g}_\alpha \cdot {\bf r}}|\psi_{m'\k'}^{(0)} \rangle \;,
\label{eq:zab}
\end{eqnarray}
\begin{eqnarray}
M_{\alpha,m\k,m'\k'}^{(1,1)} &=&
\langle \psi_{m\k,\q}^{(1)}|e^{-i{\bf g}_\alpha \cdot {\bf r}}|\psi_{m'\k',\q}^{(1)} \rangle
\nonumber\\ &&+
\langle \psi_{m\k,-\q}^{(1)}|e^{-i{\bf g}_\alpha \cdot {\bf r}} |\psi_{m'\k',-\q}^{(1)}\rangle \;,
\label{eq:zbb}
\end{eqnarray}
and $Q_\alpha=[M_\alpha^{(0,0)}]^{-1}$ where
\begin{equation}
M_{\alpha,m\k,m'\k'}^{(0,0)} =
\langle \psi_{m\k}^{(0)}|e^{-i{\bf g}_\alpha \cdot {\bf r}}|\psi_{m'\k'}^{(0)} \rangle
\;.
\label{eq:zaa}
\end{equation}

The transformation properties of the zero- and first-order wavefunctions
under translations, given by Eqs.~(\ref{eq:kbc}) and (\ref{eq:kqbc}),
impose sharp constraints upon which of the terms
in Eqs.~(\ref{eq:zba}-\ref{eq:zaa}) can be non-zero.  For example,
for $M_\alpha^{(1,0)}$ in Eq.~(\ref{eq:zba}), the term $\langle
\psi_{m\k}^{(0)}|e^{-i{\bf g}_\alpha \cdot {\bf r}} |\psi_{m'\k',\q}^{(1)}
\rangle$ is only non-zero if $\k=\k'+\q-{\bf g}_\alpha$.  Similarly,
$Q_{\alpha,m\k,m'\k'}$ is only non-zero if $\k=\k'+{\bf g}_\alpha$.
In practice, we define primitive-cell-periodic functions
\begin{equation}
u_{m\k}^{(0)}(\rr)=e^{-i\k\cdot\rr}\psi_{m\k}^{(0)}(\rr)
\end{equation}
and
\begin{equation}
u_{m\k,\q}^{(1)}(\rr)=e^{-i(\k+\q)\cdot\rr}\psi_{m\k,\q}^{(1)}(\rr)
\end{equation}
so that
\begin{eqnarray}
M_{\alpha,m\k,m'\k'}^{(1,0)} &=& S_{\k\k',mm'}^{(1,0)} \,
\delta_{\k,\k'+\q-\g_\alpha} \;,
\nonumber\\
M_{\alpha,m\k,m'\k'}^{(0,1)} &=& S_{\k\k',mm'}^{(0,1)} \,
\delta_{\k,\k'-\q-\g_\alpha} \;,
\nonumber\\
M_{\alpha,m\k,m'\k'}^{(1,1)} &=& S_{\k\k',mm'}^{(1,1)} \,
\delta_{\k,\k'-\g_\alpha} \;,
\nonumber\\
M_{\alpha,m\k,m'\k'}^{(0,0)} &=& S_{\k\k',mm'}^{(0,0)} \,
\delta_{\k,\k'-\g_\alpha} \;,
\label{eq:deltas}
\end{eqnarray}
where
\begin{eqnarray}
S_{\k\k',mm'}^{(1,0)} &=&
\langle u_{m\k}^{(0)}|u_{m'\k',\q}^{(1)} \rangle +
\langle u_{m\k,-\q}^{(1)}|u_{m'\k'}^{(0)} \rangle \;,
\nonumber\\
S_{\k\k',mm'}^{(0,1)} &=&
\langle u_{m\k}^{(0)}|u_{m'\k',-\q}^{(1)}\rangle +
\langle u_{m\k,\q}^{(1)}|u_{m'\k'}^{(0)} \rangle \;,
\nonumber\\
S_{\k\k',mm'}^{(1,1)} &=&
\langle u_{m\k,\q}^{(1)}|u_{m'\k',\q}^{(1)} \rangle +
\langle u_{m\k,-\q}^{(1)}|u_{m'\k',-\q}^{(1)}\rangle \;,
\nonumber\\
S_{\k\k',mm'}^{(0,0)} &=&
\langle u_{m\k}^{(0)}|u_{m'\k',\q}^{(0)} \rangle
\end{eqnarray}
(subscript $\alpha$ is now implicit).
Defining $Q_{\k'\k}=[S_{\k\k'}^{(0,0)}]^{-1}$ and taking
into account the constraints on k-points embodied in the delta functions
in Eq.~(\ref{eq:deltas}), the two terms in Eq.~(\ref{eq:logdeta}) become
\begin{equation}
\textrm{Tr}[M_\alpha^{(1,1)}Q_\alpha]=\sum_\k
   \textrm{Tr}[S^{(1,1)}_{\k,\k+\g} Q_{\k+\g,\k}]
\end{equation}
and
\begin{eqnarray}
&&\textrm{Tr}[M_\alpha^{(1,0)}Q_\alpha M_\alpha^{(0,1)}Q_\alpha]=
   \sum_\k \textrm{Tr}[S^{(1,0)}_{\k,\k+\g-\q}
\nonumber\\
&&\qquad
 \times Q_{\k+\g-\q,\k-\q} S^{(0,1)}_{\k-\q,\k+\g} Q_{\k+\g,\k}] \;.
\end{eqnarray}
In these equations, the trace on the left-hand side is over
all occupied states of the supercell, while on the right-hand
side it is over bands of the primitive cell.  These are the
terms that appear in Eq.~(\ref{eq:loops}) in the main text, and that
determine the pattern of k-point loops illustrated in
Fig.~\ref{fig:kpoint}.


\end{document}